\documentclass[english,aps,prd,nofootinbib,superscriptaddress,12pt]{revtex4-1}

\usepackage{graphicx} 
\usepackage{color}
\usepackage[T1]{fontenc}
\usepackage{lmodern}
\usepackage{caption}
\usepackage{multirow}
\usepackage{floatrow}

\usepackage{subcaption}
\usepackage{hyperref}
\hypersetup{colorlinks,linkcolor={blue},citecolor={blue},urlcolor={blue}}

\usepackage{mathtools}
\usepackage{physics}
\usepackage{amsmath}
\usepackage{amssymb}
\usepackage{slashed}
\usepackage{lineno}

\begin{document}


\title{Elastic neutrino-electron scattering perspectives at nuclear reactors}

\author{Luis~A.~Delgadillo} \email{ldelgadillof@ihep.ac.cn}
\affiliation{Institute of High Energy Physics, Chinese Academy of Sciences, Beijing 100049, China}
\affiliation{Kaiping Neutrino Research Center, Guangdong 529386, China}  
\author{Qishan
 Liu} \email{liuqs@ihep.ac.cn}
\affiliation{Institute of High Energy Physics, Chinese Academy of Sciences, Beijing 100049, China}  
\affiliation{China Center of Advanced Science and Technology, Beijing 100190, China} 
\author{Randhir~Singh} \email{singhrandhir@ihep.ac.cn}
\affiliation{Institute of High Energy Physics, Chinese Academy of Sciences, Beijing 100049, China}  
\affiliation{Kaiping Neutrino Research Center, Guangdong 529386, China}   
\date{\today}

\begin{abstract}
\noindent 
The determination of the weak mixing angle, $\sin^2\theta_W$, at low momentum transfers remains a powerful test of the Standard Model and its potential new physics extensions. In this paper, we explore some physics opportunities at present and future reactor neutrino experiments through elastic neutrino-electron scattering (E$\nu$ES). We assess the expected sensitivity to the weak mixing angle considering the CLOUD, TAO, and DANSS experimental configurations. We find that both CLOUD and TAO may achieve a precision that surpasses the current global fit from reactor experiments, while DANSS alone is expected to surpass the benchmark precision set by TEXONO measurement of the weak mixing angle. Additionally, we derive projected upper limits for the non-standard neutrino interactions (NSI), effective neutrino magnetic moment ($\mu_\nu$) and translate these into constraints on the neutrino transition magnetic moments ($\Lambda_i$). Our results demonstrate the physics potential of the E$\nu$ES channel at current and upcoming reactor-based neutrino experiments.
\end{abstract}

\maketitle

\section{Introduction}
\label{sec:intro}
Present neutrino oscillation datasets have determined two neutrino mass squared differences, $\Delta m_{21}^{2}$ and $\Delta m_{31}^{2}$, the corresponding mixing angles $\theta_{ij}$ (with $ij={12,13,23}$)~\cite{ParticleDataGroup:2024cfk,JUNO:2025gmd}, and provided hints on the value of the Dirac CP-violating phase $\delta_{CP}$~\cite{NOvA:2021nfi, T2K:2023smv, T2K:2025wet}. In the establishment of the three-neutrino oscillation paradigm, reactor neutrino experiments such as Daya Bay~\cite{DayaBay:2012fng}, RENO~\cite{RENO:2012mkc}, and Double Chooz~\cite{DoubleChooz:2012gmf} played a significant role by measuring a non-zero mixing angle ($\theta_{13}$). Nevertheless, other interesting physical studies can be investigated in reactor neutrino experiments. 

The elastic neutrino-electron scattering (E$\nu$ES) can be employed to further reduce neutrino flux uncertainties, as demonstrated by the MINER$\nu$A collaboration~\cite{MINERvA:2022vmb}. Therefore, at reactor-based neutrino experiments, in addition to the main inverse beta decay (IBD) channel, the E$\nu$ES process and the coherent elastic neutrino-nucleus scattering channel~\cite{Liao:2023kyy} could also be used to improve neutrino flux predictions below the IBD threshold. Besides, reactor neutrino experiments with neutrino energies of the order of a few MeV can probe the weak mixing angle~\cite{Conrad:2004gw} and the electromagnetic properties of neutrinos~\cite{Giunti:2008ve} through the elastic scattering of antineutrinos off electrons~\cite{Broggini:2012df, Giunti:2014ixa, Giunti:2024gec}. 
The interaction between low-energy neutrinos and electrons may potentially involve new physics beyond the Standard Model (BSM). Such effects can be parameterized in a model-independent way by neutrino non-standard interactions (NSI)~\cite{Bischer:2018zcz,Bischer:2019ttk,Han:2020pff,Dutta:2017nht,Miranda:2015dra}, which would induce an effective shift in the measured value of the weak mixing angle. Therefore, precision measurements of $\sin^2\theta_W$ at reactors also serve as sensitive probes for NSI.

The weak mixing angle, or Weinberg angle ($\sin^2 \theta_W$), is an important parameter within the SM~\cite{Weinberg:1967tq}. It has been precisely measured at high energies by the LEP collider~\cite{ALEPH:2005ab}, as well as the NuTeV collaboration~\cite{NuTeV:2001whx}. However, only a few measurements or projected experiments of this parameter exist near the MeV scale; see, e.g., Refs.~\cite{Giunti:2014ixa, Erler:2017knj,Yip:2025dxe,AtzoriCorona:2022qrf}. Regarding reactor neutrino experiments, a measurement of the weak mixing angle via the E$\nu$ES process was determined by the TEXONO collaboration~\cite{TEXONO:2009knm}. 
Besides, measurements from atomic parity violation~\cite{Antypas:2018mxf}, provide the most precise determination of the weak mixing angle at low energies. For a global assessment of the weak mixing angle from low-energy neutrino experiments we refer the reader to Refs.~\cite{Canas:2016vxp, Barranco:2007ea}.

The concept of the neutrino and its electromagnetic properties, including the possibility of a magnetic moment ($\mu_\nu$), originated with Pauli's hypothesis of the particle's existence. However, systematic theoretical investigations started much later, after models incorporating right-handed neutrinos demonstrated that a massive neutrino could generally possess a magnetic moment~\cite{Fujikawa:1980yx, Shrock:1982sc}. The possibility for E$\nu$ES due to neutrino magnetic moment was first considered by Carlson and Oppenheimer~\cite{Carlson:1932rk} and the cross section of this process was calculated later~\cite{Domogatsky:1970zvt}. For instance, the neutrino magnetic moment contribution to the elastic scattering process flips the neutrino helicity. If neutrinos are Dirac particles, this process transforms active left-handed neutrinos into sterile right-handed neutrinos~\cite{PhysRevD.36.2278}. In the SM, neutrinos are massless, and they cannot have a magnetic or electric dipole moment. However, massive (Dirac or Majorana) neutrinos can develop a magnetic moment at the one-loop level~\cite{Fujikawa:1980yx}.

Experimental searches for neutrino magnetic moments have been extensively investigated via the E$\nu$ES process. Different solar, accelerator and reactor measurements which rely on the elastic scattering process, have put constraints on the effective neutrino magnetic moment~\cite{Vidyakin:1992nf, Derbin:1993wy, Allen:1992qe, LSND:2001akn, TEXONO:2006xds, MUNU:2005xnz, Borexino:2008dzn, Balantekin:2004tk, Super-Kamiokande:2004wqk, DONUT:2001zvi, Beda:2012zz}, and bounds are approaching the level where the SM extension with massive neutrinos predicts a non-zero but small neutrino magnetic moment~\cite{Giunti:2008ve, Broggini:2012df, Giunti:2014ixa, Bernabeu:2000hf, Bernabeu:2002pd, AtzoriCorona:2025xgj}. As far as reactor neutrino experiments are concerned, the most stringent limit on the effective electron-neutrino magnetic moment $\mu_{\nu_e}$ has been obtained by the GEMMA collaboration~\cite{Beda:2012zz}. Nevertheless, while $\mu_\nu$ serves as a useful phenomenological benchmark for experimental comparison, it is not the foundational theoretical entity. The fundamental quantities are, in fact, the neutrino transition magnetic moments, $\Lambda_i$. These intrinsic matrix elements connect different neutrino mass eigenstates and arise directly from models beyond the Standard Model. The experimentally inferred effective moment $\mu_\nu$, is thus a composite measure, constructed from the fundamental quantities, $\Lambda_i$ as projections onto a specific experimental and flavor context~\cite{Ternes:2025lqh}. 

The electromagnetic properties of neutrinos are of critical importance, as measuring these properties could reveal whether neutrinos are Dirac or Majorana particles~\cite{Schechter:1981hw}, and may provide evidence of BSM physics~\cite{Bell:2005kz}. However, despite extensive research, no experimental evidence has yet confirmed non-zero neutrino electromagnetic properties.

Current and upcoming reactor neutrino experiments that could determine the weak mixing angle at MeV energies, as well as probe the effective neutrino magnetic moment include DANSS~\cite{Alekseev:2016llm}, JUNO-TAO (henceforth TAO)~\cite{JUNO:2020ijm} and CLOUD~\cite{NavasNicolas:2024ixn}. The DANSS (Detector of the reactor AntiNeutrino based on Solid Scintillator) experiment is a solid scintillator neutrino detector located in Udomlya, Russia. Constructed as a compact neutrino spectrometer, it is capable of being positioned immediately next to a reactor core. Alternatively, TAO (Taishan Antineutrino Observatory) is a liquid scintillator neutrino experiment based in Guangdong province in China and currently is in its commissioning phase. In addition, the CLOUD (Chooz LiquidO Ultra near Detector) proposal, is the third generation (after CHOOZ and Double Chooz experiments) of neutrino fundamental studies at the Chooz nuclear power plant based in France, envisioned as a new frontier in neutrino physics employing the novel LiquidO technology~\cite{LiquidO:2019mxd}. For an overview of the current status and future particle physics potential using reactor antineutrinos we refer the reader to Ref.~\cite{Akindele:2024nzu}.

This paper is organized as follows. Section~\ref{sec:framework} presents the formalism of elastic neutrino--electron scattering. Furthermore, the experimental configurations considered in this work are detailed in Section~\ref{sec:setup}. Section~\ref{method} describes the general methodology employed to obtain the expected signal and background events for each setup. In Section~\ref{sec:results}, we present and discuss our results of the expected sensitivities, and finally, Section \ref{sec:conclusions} provides the conclusions and future assessments in this direction.

\section{Theoretical framework}
\label{sec:framework}

\subsection{Elastic neutrino--electron scattering}
\label{subsec:enues}
The differential cross section for the elastic neutrino--electron scattering process (E$\nu$ES) at tree-level in the Standard Model (SM) reads~\cite{Vogel:1989iv}
\begin{equation}
\label{xsc}
    \frac{d\sigma(T_e, E_\nu)}{dT_e} = \frac{G_F^2 m_e}{2 \pi} \Big[ (g_V-g_A)^2 +(g_V+g_A)^2 \Big(1-\frac{T_e}{E_\nu} \Big)^2-\big(g_V^2-g_A^2 \big) \frac{m_eT_e}{E_\nu^2} \Big]\;,
\end{equation}
here, $T_e$ and $m_e$ are the recoil kinetic energy and mass of the electron, $G_F$ is the Fermi constant, $E_\nu$ is the neutrino energy, and $g_V= 2\sin^2 \theta_{W}+1/2,~g_A=+1/2$ are the corresponding vector ($V$) and axial ($A$) couplings, with $\theta_{W}$ the weak mixing angle. Throughout this work, as benchmark value of the weak mixing angle, we consider the SM best-fit at low energy $\sin^2 \theta_{W} = 0.2385$~\cite{Tiesinga:2021myr}. However, aside from the contributions of the effective neutrino magnetic moment and neutrino charge radius~\footnote{The effective neutrino charge radius contribution is included through a shift of the weak mixing angle, $\sin^2 \theta_{W} \rightarrow \sin^2 \theta_{W}\big(1+\frac{M_W^2}{3} \langle r^2_\nu \rangle\big)$ ~\cite{Cadeddu:2018dux}.}, in this analysis, we will not consider other radiative corrections to the E$\nu$ES cross section~\cite{Tomalak:2019ibg, Brdar:2024lud, Huang:2024rfb}, as their impact on the projected sensitivities is negligible for our assumed systematic uncertainties (see Tab.~\ref{tab:sys})~\cite{Bahcall:1995mm}. 

\subsection{Neutrino magnetic moments}
In the simplest extension of the SM with three massive Dirac neutrinos, the diagonal magnetic moments are significantly suppressed by the smallness of neutrino masses~\cite{Giunti:2024gec}
\begin{equation}
    \mu_{\nu}^{\text{D}} = \frac{3e G_F m_i}{8 \sqrt{2}\pi^2} \simeq 3\times10^{-20} \big(\frac{m_i}{0.1~\text{eV}} \big) \mu_{\text{B}}~,
\end{equation}
here, $e$ is the elementary charge, $m_i$ is the neutrino mass, and $\mu_{\text{B}} \equiv e /(2 m_e)$ is the corresponding Bohr magneton. 
For instance, the neutrino magnetic moment ($\mu_{\nu_\ell}$) contributes to the elastic neutrino-electron scattering cross section as~\cite{Vogel:1989iv}
\begin{equation}
\label{mmxsec}
    \frac{d \sigma (\mu_{\nu_\ell})}{dT_e}  = \frac{\pi \alpha^2 }{m_e^2} \abs{\frac{\mu_{\nu_\ell}^2}{\mu_{\rm B}^2}} \Bigg( \frac{1}{T_e} -\frac{1}{E_\nu}\Bigg)~,
\end{equation}
$\alpha$ being the fine-structure constant. Due to its $1/T_e$ functional dependence on the cross section, the sensitivity to the magnetic moment is enhanced for low values of $T_e$ at given $E_\nu$.

However, in order to provide an equivalent comparison of neutrino magnetic moments with respect to solar, reactor and other measurements, the fundamental quantities to contrast are the corresponding transition magnetic moments~\cite{Ternes:2025lqh}. For instance, in the case of Majorana neutrino magnetic moments, the effective Hamiltonian is given as~\cite{Schechter:1981hw, Canas:2015yoa}
\begin{equation}
H^{\rm{M} }_{\rm{EM}} = -\frac{1}{4}\nu^{T}_{L}\mathcal{C}^{-1}~ \lambda_{\rm{M}}~\sigma^{\alpha\beta}\nu_{L}F_{\alpha\beta} + \rm{h.c.}\;,
\end{equation}
where $\lambda_{\rm{M}} = a - ib$ is an antisymmetric complex matrix, $\lambda_{\rm{M}}^{\alpha\beta}=-\lambda_{\rm{M}}^{\beta\alpha}$, so that $a^{T}=-a$ and $b^{T}=-b$ are imaginary. Hence, three complex or six real parameters are necessary to describe the Majorana neutrino case. On the other hand, for the case of Dirac neutrino magnetic moments, the corresponding Hamiltonian reads~\cite{Grimus:2000tq}
\begin{equation}
H^{\rm{D}}_{\rm{EM}} = \frac{1}{2}\bar{\nu}_{R}~\lambda_{\rm{D}}~\sigma^{\alpha\beta}\nu_{L}F_{\alpha\beta} + \rm{h.c.}\;,
\end{equation}
with $\lambda_{\rm{D}} = a + ib$ being an arbitrary $3\times3$ complex matrix. Besides, hermiticity of $H^{\rm{D}}_{\rm{EM}}$ implies $a=a^{\dag}$, and $b=b^{\dag}$. From a phenomenological point of view, the Dirac neutrino magnetic moment is described as $ \lambda_{\rm{D}} = a + ib$ ($\tilde{\lambda}_{\rm{D}}$) in the flavor (or mass) basis, while for the Majorana case, $\lambda_{\rm{M}}$ can be parameterized as~\cite{Canas:2015yoa, Grimus:2002vb, Ternes:2025lqh}
\begin{equation}
\lambda_{\rm{M}} = \left(\begin{array}{ccc}
0 & \Lambda_{\tau} & -\Lambda_{\mu} \\ -\Lambda_{\tau} & 0 & \Lambda_{e} \\ \Lambda_{\mu} & -\Lambda_{e} & 0 \end{array}\right) , \qquad
\tilde{\lambda}_{\rm{M}} = \left(\begin{array}{ccc} 0 & \Lambda_{3} & -\Lambda_{2} \\
-\Lambda_{3} & 0 & \Lambda_{1} \\
\Lambda_{2} & -\Lambda_{1} & 0 \end{array}\right)\;,
\label{Eq:nmm-matrix}
\end{equation}
being the transition magnetic moments $\Lambda_\alpha$ and $\Lambda_j$: $\Lambda_{\alpha}=|\Lambda_{\alpha}|e^{i \omega_{\alpha}}$, and $\Lambda_{j}=|\Lambda_{j}|e^{i\omega_{j}}$. 

In this analysis, we consider the case of Majorana transition moments in the mass basis ($\tilde{\lambda}_{\rm{M}}$), assessing the impact of one parameter at a time and fixing all other transition magnetic moment parameters and phases ($\omega_j$) to zero, see, e.g., Ref.~\cite{Canas:2015yoa}. For simplicity, we will omit the superscript (M) in $\mu_\nu^{\rm{M}}$ referring to Majorana neutrinos. At short baseline reactor neutrino experiments such as those considered in this analysis, the relationship between the effective magnetic moment and transition magnetic moments is given by~\cite{Canas:2015yoa, Grimus:2000tq, Grimus:2002vb}

\begin{equation}
\begin{split}
\mu_\nu^{2} = & |\Lambda|^{2} - s^{2}_{12}c^{2}_{13}|\Lambda_{2}|^{2} - c^{2}_{12}c^{2}_{13}|\Lambda_{1}|^{2} - s^{2}_{13}|\Lambda_{3}|^{2}\\
-& 2s_{12}c_{12}c^{2}_{13}|\Lambda_{1}||\Lambda_{2}|\cos\delta_{12} - 2c_{12}c_{13}s_{13} 
|\Lambda_{1}||\Lambda_{3}|\cos\delta_{13}\\
-& 2s_{12}c_{13}s_{13}|\Lambda_{2}||\Lambda_{3}|\cos\delta_{23}\;,
\end{split}
\label{eq:mueff}
\end{equation}
where $c_{ij} = \cos\theta_{ij}$, $s_{ij} = \sin\theta_{ij}$ are the corresponding mixing angles in the $ij-$sector of the standard three neutrino oscillations framework, $\delta_{12}= \omega_{2}-\omega_{1}$, $\delta_{23}= \omega_{3}-\omega_{1} - \delta_{CP}$, and $\delta_{13}=
\delta_{12}-\delta_{23}$ are the additional Majorana phases, with $\delta_{CP}$ being the Dirac $CP$-violating phase of the leptonic mixing matrix. 

\subsection{Neutrino NSI}
To search for BSM, the neutrino NSI with electron for energies $\ll M_Z$ can be described by the effective Lagrangian~\cite{Bischer:2018zcz,Bischer:2019ttk,Han:2020pff,Dutta:2017nht,Miranda:2015dra}
\begin{equation}
\label{611}
\mathcal{L}_{\nu }^{NSI}=-\frac{G_{F}}{\sqrt{2}} \sum_{ \alpha, \beta=e, \mu, \tau}\left[\bar{\nu}_{\alpha} \gamma^{\mu}\left(1-\gamma^{5}\right) \nu_{\beta}\right]\left(\varepsilon_{\alpha \beta}^{e L}\left[\bar{e} \gamma_{\mu}\left(1-\gamma^{5}\right) e\right]+\varepsilon_{\alpha \beta}^{e R}\left[\bar{e} \gamma_{\mu}\left(1+\gamma^{5}\right) e\right]\right),
\end{equation}
where $\varepsilon_{\alpha \beta}^{e P}$ (with $\alpha, \beta=e, \mu, \tau$ and $P=L, R$) describe non-standard neutrino interactions, including both the non-universal terms $\varepsilon_{\alpha \alpha}^{e P}$ and the flavor-changing contributions $\varepsilon_{\alpha \beta}^{e P}$ ($\alpha \neq \beta$). These interactions can be characterized by vector couplings and axial-vector couplings. The vector couplings are represented by the spin-independent combination $\varepsilon_{\alpha \beta}^{e V}=\varepsilon_{\alpha \beta}^{e L}+\varepsilon_{\alpha \beta}^{e R}$, while the axial-vector couplings are characterized by the orthogonal spin-dependent combination $\varepsilon_{\alpha \beta}^{e A}=\varepsilon_{\alpha \beta}^{e L}-\varepsilon_{\alpha \beta}^{e R}$.
The vector and axial couplings in Eq.~\ref{xsc} would be modified to $g_V\rightarrow 2\sin^2 \theta_{W}+1/2 + \varepsilon_{\alpha \beta}^{e V}$, and $g_A\rightarrow 1/2+ \varepsilon_{\alpha \beta}^{e A}$, accordingly~\cite{Miranda:2015dra}. 

\section{Experimental configurations}
\noindent
\label{sec:setup}
In this section, we describe the main characteristics of the three reactor neutrino experiments considered in this analysis—CLOUD, TAO, and DANSS--located at short baselines of approximately 10--40 meters from the main nuclear reactor cores. These facilities provide intense, pure sources of electron antineutrinos ($\bar{\nu}_e$), enabling high-statistics measurements of neutrino interactions. While the primary detection channel in all three experiments is IBD, this work focuses on the determination of the weak mixing angle $\sin^2\theta_W$ at low momentum transfer and searches for physics BSM through the E$\nu$ES process. This channel is particularly sensitive to the effective neutrino magnetic moment, $\mu_\nu$, due to its energy threshold. Complementing these E$\nu$ES measurements, the experiments will provide precise measurements of the reactor antineutrino energy spectrum via the IBD channel. 

\subsection{CLOUD}
\label{subsec:cloud}
The proposed Super-Chooz reactor neutrino experiment is envisioned for the site of the former Double Chooz experiment~\cite{DoubleChooz:2012gmf,CLOUD:2024,SuperChooz:2022,AntiMatter-OTech:2025}. As a preliminary assessment for this project, the CLOUD (Chooz LiquidO Ultra-near Detector) experiment has been conceived. CLOUD will employ a novel opaque scintillator technique known as LiquidO~\cite{LiquidO:2019mxd}. This technology builds upon existing scintillator detector designs, utilizing components such as organic scintillator materials, wavelength-shifting fibers, and photo-sensors~\cite{LiquidO:2025qia}. The CLOUD experiment will feature a LiquidO detector with a mass of $\leq$10 tons, situated approximately 35 meters from the Chooz B2 reactor core. With the reactor producing a thermal power of 4.25 GW, CLOUD is projected to detect of the order of $10^4$ antineutrinos per day.

Besides, CLOUD experiment seeks to achieve unmatched precision in measuring the absolute reactor antineutrino flux and in determining the weak mixing angle ($\sin^2 \theta_W$) at MeV energies through the neutrino-electron elastic scattering process. It will also perform searches for BSM physics~\cite{NavasNicolas:2024ixn}. The project will be executed in phases: Phase-I will concentrate on reactor antineutrino observations, while subsequent phases (Phase-II and Phase-III) will assess the viability of detecting solar neutrinos, geoneutrinos, and other physics opportunities.

\subsection{TAO}
\label{subsec:tao}
The Taishan Antineutrino Observatory (TAO), also known as JUNO-TAO, will serve as a companion experiment to JUNO~\cite{JUNO:2020ijm,JUNO:2015sjr}, aiming to deliver a reactor antineutrino spectrum measurement with energy resolution below few percent. TAO's measurements via IBD process will offer a precise reference spectrum for JUNO analyses and act as a critical test for current nuclear databases as well as conducting searches for light sterile neutrinos~\cite{JUNO:2020ijm, Berryman:2021xsi}. The TAO detector will consist of approximately one ton (fiducial volume) of liquid scintillator, it is located just 44 meters from one of the Taishan Nuclear Power Plant’s reactor core~\cite{JUNO:2024jaw}. Its operational timeline is expected to coincide with that of JUNO. TAO IBD rates are estimated to be 1000 (2000) events per day in the fiducial volume with (without) the detection efficiency taken into account. 

TAO central detector (CD) consists of a spherical acrylic vessel which has an inner diameter of 1.8 m, it contains approximately 2.8 tons of Gadolinium-doped Liquid Scintillator (GdLS). In order to reduce background contaminants at the CD, a fiducial volume (FV) cut of 25 cm is considered, it will reduce the detector material FV to 1 ton in a radius of 65 cm. For instance, the light yield is about 4500 photons per MeV, hence, for the case of electrons, TAO is expected to achieve an energy resolution below $2\%$ at 1 MeV~\cite{JUNO:2020ijm}.

\subsection{DANSS}
\label{subsec:danss}
The DANSS detector is composed of a segmented 1-m$^3$ plastic scintillator~\cite{Alekseev:2016llm, Danilov:2024fwi}. It features scintillator strips coated with a gadolinium-infused reflective surface. DANSS is installed under the core of a 3.1 GW (thermal power) reactor at the Kalinin Nuclear Power Plant (KNPP) on a moving platform. It operates at distances ranging from 9.7 to 12.2 meters from the reactor core~\footnote{However, in this study, we consider a mean baseline at 11 meters.}. For instance, the measured IBD rate in the fiducial volume of the DANSS detector is about 5,000 events per-day~\cite{DANSS:2018fnn}, approximately five times larger than those expected at TAO. As one of its main physics goals, the DANSS experiment searched for a light sterile neutrino ($\sim$1 eV mass) by measuring the reactor antineutrino energy spectrum and its variation across multiple distances from the reactor core, it has set stringent exclusion limits in the sterile neutrino parameter space ($\sin^22\theta_{ee}$ vs. $\Delta m^2_{41}$)~\cite{DANSS:2018fnn}, ruling out a significant portion of the parameter region previously suggested by the reactor antineutrino anomaly~\cite{Mention:2011rk} and Gallium anomaly~\cite{Kostensalo:2019vmv} at $>$95\% CL, particularly for $\Delta m^2_{41} \sim 1.3$ eV$^2$. 

Recently, the DANSS collaboration performed a search for oscillation patterns predicted in some extra dimension models, it has found no evidence, and thus, ruled out extra dimension explanations for the reactor antineutrino anomaly~\cite{Alekseev:2025tpq}. DANSS is planning an upgrade after which it will have an energy resolution of around 12$\%$ at 1 MeV~\cite{Svirida:2024jfb}.

\section{Methodology}
\label{method}
\subsection{Signal events}
As far as reactor antineutrino fluxes is concerned, we construct the spectra for ${}^{235}\mathrm{U}$, ${}^{238}\mathrm{U}$, ${}^{239}\mathrm{Pu}$, and ${}^{241}\mathrm{Pu}$ using a fifth-order polynomial parametrization by Huber and Mueller (HM) model~\cite{Huber:2011wv, Mueller:2011nm} with fission fractions as given in Table~\ref{tab:1}. However, below the IBD threshold, both conversion and summation models~\cite{Estienne:2019ujo, Vogel:1989iv} are considered. 

\begin{table}[h]
    \centering
    \caption{Reactor fuel fractions considered in this study.} 
    \begin{tabular}{c c c c c}
        \hline
        Isotope & ${}^{235}$U & ${}^{238}$U & ${}^{239}$Pu & ${}^{241}$Pu \\
        \hline \hline
        CLOUD~\cite{DoubleChooz:2011ymz} & 0.488 & 0.087 & 0.359 & 0.067 \\
        TAO~\cite{JUNO:2024jaw} & 0.561 & 0.076 & 0.307 & 0.056 \\
        DANSS~\cite{Kopeikin:2012zz} & 0.56 & 0.07 & 0.31 & 0.06 \\
        \hline
    \end{tabular}
    \label{tab:1}
\end{table}

In this analysis, we established a total of 30, 60, and 24 energy bins for the CLOUD, TAO, and DANSS detectors, respectively. These bins were uniformly distributed over the region of interest (ROI) in electron recoil energy $T_e$, which ranged from $T_e^{\rm min} = 1.0$ MeV to $T_e^{\rm max} = 7$ MeV. Furthermore, the corresponding neutrino energy ranges were set from $E_{\nu}^{\text{min}} = \frac{1}{2} \big(T_e^{\text{min}} +\sqrt{(T_e^{\text{min}})^2 +2 m_e T_e^{\text{min}}}\big)\simeq1.211$ MeV to $E_{\nu}^{\text{max}} = 10$ MeV. For this study, the assumed values of the detector fiducial volumes are 10 tons for CLOUD and one ton for both the TAO and DANSS configurations. Additionally, for all setups, we consider a duty factor of $\tau = 11/12$ and a signal detection efficiency of 90\% in the corresponding ROI. For instance, the assumed threshold in this analysis is lower than those of previous reactor neutrino experiments, such as the one reported by the TEXONO collaboration, $T_e^{\rm min} \simeq 3.0$ MeV~\cite{TEXONO:2009knm}. The lower threshold considered in this study could improve the sensitivity to the neutrino magnetic moment due to its low-energy enhancement. Furthermore, the assumed detection efficiency of 90\% is consistent with values reported in a similar experimental configuration~\cite{JUNO:2024jaw}.

The predicted differential event rates including energy resolution effects can be computed as (see, e.g., Refs.~\cite{Canas:2016vxp, Brdar:2024lud})
\begin{equation}
    \frac{dN}{dT_e^{\prime}} = \mathcal{N}~\epsilon \int_{T_e^{\rm min}}^{T_e^{\rm max}} dT_e ~R(T_e,~T_e^{\prime})  \int_{E_{\nu}^{\text{min}}}^{E_{\nu}^{\text{max}}} dE_{\nu}~\Phi_{\nu}(E_\nu) \frac{d \sigma (T_e, E_\nu)}{dT_e}~,
\end{equation}
here, $\mathcal{N}$ includes the detector volume and time exposure (10-years) including duty factor ($\tau = 11/12$), $\epsilon$ is the signal detection efficiency ($90\%$), $d \sigma (T_e, E_\nu) / dT_e$ is the E$\nu$ES differential cross section, $\Phi_{\nu} (E_\nu)$ is the electron antineutrino flux, and $T_e$ is true kinetic energy of the electron. Besides, detector response which relates the true ($T_e$) and reconstructed electron energies ($T_e^{\prime}$), is parameterized via Gaussian smearing
\begin{equation}
\label{eq:eres}
  R(T_e,\,T_e^{\prime})  = \frac{1}{ \sqrt{2 \pi} \sigma} \exp{\frac{-(T_e - T_e^{\prime})^2}{2 \sigma^2}}\,,
\end{equation}
with corresponding resolution ($\sigma/T_e$) for each setup as given in Table~\ref{tab:2}. In order to obtain the expected event rates, we integrate the reconstructed event rates $dN/dT_e^{\prime}$ across a given energy bin ($i$). 

\begin{table}[h]
    \centering
    \caption{Experimental configurations considered in this work.} \label{tab:2}
    \begin{tabular}{c c c c c}
        \hline
        Experiment & Energy resolution  & Bin-width & Baseline & Exposure \\
        \hline \hline
        CLOUD~\cite{LiquidO:2019mxd} & $0.05  \sqrt{T_e /\rm{MeV}}$ & 0.2 MeV & $L= 35$ m & 10 ton$\times$year\\
        TAO~\cite{JUNO:2020ijm} & $0.02  \sqrt{T_e/\rm{MeV}}$  & 0.1 MeV& $L= 44$ m& 1 ton$\times$year \\
        DANSS~\cite{Svirida:2024jfb} & $0.12  \sqrt{T_e/\rm{MeV}}$ & 0.25 MeV & $L= 11$ m & 1 ton$\times$year\\
        \hline
    \end{tabular}
\end{table}

\subsection{Backgrounds}
\label{subsec:bckg}
\begin{figure}[H]
\includegraphics[scale=0.65]{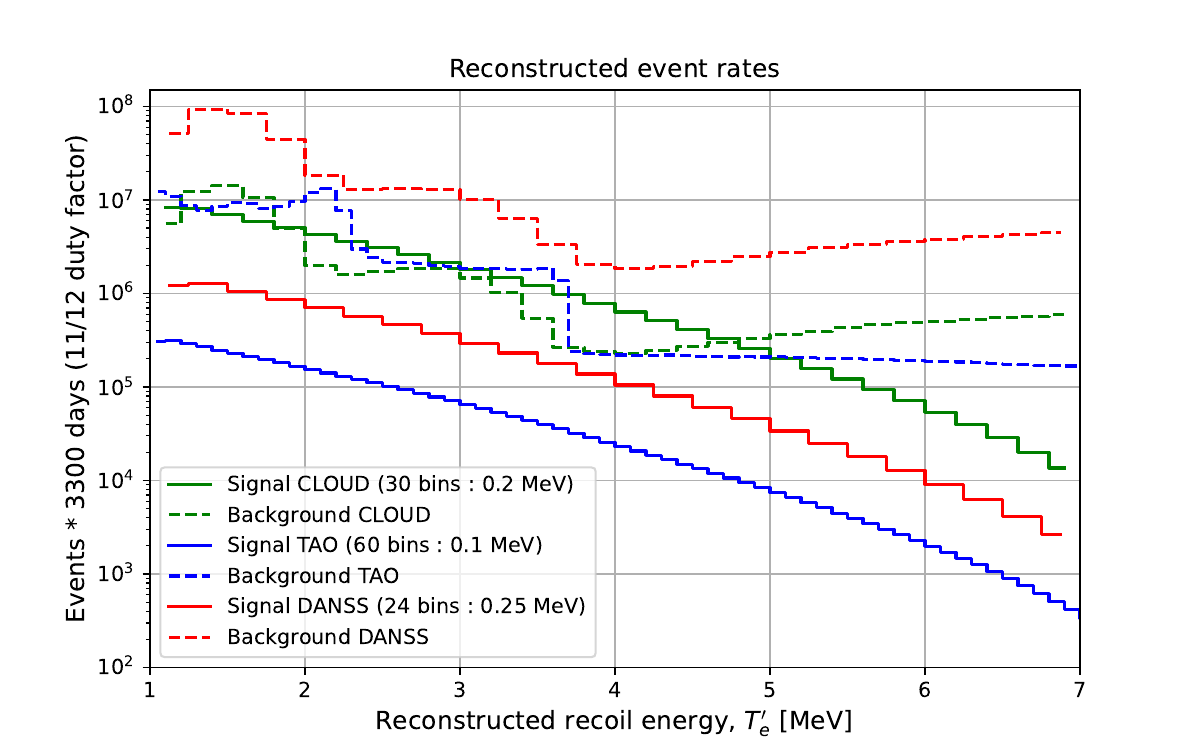}
		 \caption{Expected reconstructed signal and background event rates for the corresponding reactor setups. Signal events are shown in solid lines for the CLOUD (green), TAO (blue), and DANSS (red) configurations. Similarly, expected background events are shown in dashed lines. All projections assume a 10‑year exposure with a duty factor of 11/12, using target masses of 10 ton for CLOUD and 1 ton for both TAO and DANSS}. See the text for details.
  \label{fig:f0}
\end{figure}

As far as the CLOUD and DANSS experiments are concerned, we consider the expected backgrounds from Ref.~\cite{Conrad:2004gw}. The background model consists of singles from the $^{40}$K isotope and contributions from $^{238}$U and $^{232}$Th chains. In addition, low energy neutrons and muon-induced spallation contaminants are included. For the CLOUD experiment we consider that the LiquidO technology~\cite{LiquidO:2019mxd} can reduce the background contamination down to 300 meter of water equivalent (m.w.e.). As for the DANSS setup, the same background model is assumed and rescaled to 50 m.w.e.~\cite{Alekseev:2016llm}. Regarding TAO, background contaminants were incorporated as those given in Ref.~\cite{JUNO:2024jaw} (see, e.g., Fig.~4 and Table~4 therein). The dominant sources of background in the experiment are twofold. First, there are correlated backgrounds induced by cosmic-ray muons, which produce fast neutrons and unstable isotopes such as \(^{9}\text{Li}\) and \(^{8}\text{He}\). Second, there is an accidental background arising from the random coincidence between a single neutron capture event and signals from natural radioactivity. Besides, while the reactor is operating, the dominant background is the IBD. However, this background can be effectively mitigated, as the Gd-doped detector's tagging capabilities allow IBD events to be identified through their prompt and delayed coincidence signature~\cite{JUNO:2020ijm}. Hence, in this work, we consider residual IBD background rates to be negligible.
The resulting expected signal and background events for each experimental configuration are displayed in Fig.~\ref{fig:f0}.
Fig.~\ref{fig:f0} shows the reconstructed signal and background event rates for all configurations. The signal events for CLOUD (green), TAO (blue), and DANSS (red) are shown in solid lines, while the expected backgrounds are represented by dashed lines. Owing to its greater detector mass exposure and superior background rejection, CLOUD is projected to achieve the highest signal-to-background ratio (S/B). DANSS, however, has a planned upgrade that may further reduce its background levels. In contrast to CLOUD and DANSS, TAO's performance on S/B is limited due to its greater distance from the reactor core as well as less effective background shielding.

\section{Results}
\label{sec:results}
In this section, we present the methodology of our $\chi^2$ analysis definition to assess the expected sensitivities at the aforementioned experimental configurations. In order to compute the sensitivity to the weak mixing angle and effective neutrino magnetic moment, we perform a binned $\chi^2$ analysis given by
\begin{equation}
\label{eq:chi2}
\begin{aligned}
    & \chi^2 = \sum^{ N_{ \rm{bin} } }_{i=1} \frac{\Big(S_i+ B_i-(1+\alpha)\times\eta~P_i(\Omega)-(1+\beta) B_i\Big)^2}{S_i+B_i+\sigma_S^2S_i^2 +\sigma_B^2B_i^2 } +\left(\frac{\alpha}{\sigma_\alpha}\right)^2+\left(\frac{\beta}{\sigma_\beta}\right)^2+ \left(\frac{\eta-1}{\sigma_\eta}\right)^2\;,
\end{aligned}
\end{equation}
where $S_i=P_i(\sin^2\theta_W = 0.2385,~\mu_\nu = 0)$ are the simulated signal events, $P_i(\Omega =\sin^2\theta_W~\rm{or}~\mu_\nu)$ are the predicted signal events as a function of either the weak mixing angle $\sin^2\theta_W$, or the effective neutrino magnetic moment $\mu_\nu$, and $B_i$ are the corresponding background events. 

The systematic uncertainties in this analysis are set to benchmark values representative of the performance achievable in current and future reactor neutrino experiments. We consider two scenarios: a conservative scenario, with a signal shape uncertainty of $\sigma_S = 5\%$ and a background shape uncertainty of $\sigma_B = 1\%$,  comparable to the precision attained by existing experiments such as STEREO~\cite{STEREO:2019ztb,STEREO:2020hup} and PROSPECT~\cite{PROSPECT:2020sxr, PROSPECT:2024gps}; and an optimistic scenario with $\sigma_S = 3\%$ and $\sigma_B = 0.2\%$, reflecting the improved systematic control anticipated for detectors such as TAO~\cite{JUNO:2020ijm, JUNO:2024jaw}.
We assume $\sigma_\alpha=$ 5\%, and $\sigma_\beta=$ 10\%, for signal and background normalization systematics, respectively. The nuisance parameter $\eta$ accounts for the energy scale uncertainty and we consider $\sigma_\eta=$ 1\%. Additionally, $\sigma_S$ is the uncorrelated bin-to-bin (b2b) signal shape uncertainty, while $\sigma_B$ refers to the background b2b uncorrelated shape uncertainty. We have minimized over the $\alpha$, $\beta$, and $\eta$ nuisance parameters accordingly. For all the experimental configurations, we assume the same values of systematic uncertainties; those values are summarized in Table~\ref{tab:sys}.

\begin{table}[H]
\caption{Benchmark systematic uncertainties assumed in this analysis.}
\begin{center}
\begin{tabular}{c|c}
	\hline \hline
	\multicolumn{2}{c}{Systematic error [$\%$]}\\
    \hline \hline
    \multicolumn{2}{c}{Signal systematics~~~~~~~~~~Background systematics}\\
    \hline
   Normalization: $\sigma_{\alpha}=5.0$  ~~&~~ Normalization: $\sigma_{\beta}=10$  \\
    \hline 
    Shape: $\sigma_{S}=3.0$ or $5.0$  & Shape: $\sigma_B=0.2$ or $1.0$  \\
    \hline 
    ~Energy scale: $\sigma_{\eta}=1.0$ ~& \\
	\hline \hline
\end{tabular}
\end{center}
\label{tab:sys}
\end{table}

Our projected sensitivities were calculated based on the $\Delta \chi^2 = \chi^2 - \chi ^2_{\text{min}}$ distribution (see, e.g., Ref.~\cite{Blennow:2013oma}); we scan the test parameter $\Omega= \sin^2 \theta_W~\rm{or}~\mu_\nu$, and map the $\Delta \chi^2$ to the corresponding confidence levels (CL) of the $\chi^2$ distribution considering Wilks' Theorem~\cite{Wilks:1938dza}.

\subsection{Weak mixing angle and effective neutrino magnetic moment}
\begin{figure}[H]
\begin{subfigure}[h]{0.495\textwidth}
			\caption{}
			\label{fb2bs1}
\includegraphics[width=\textwidth]{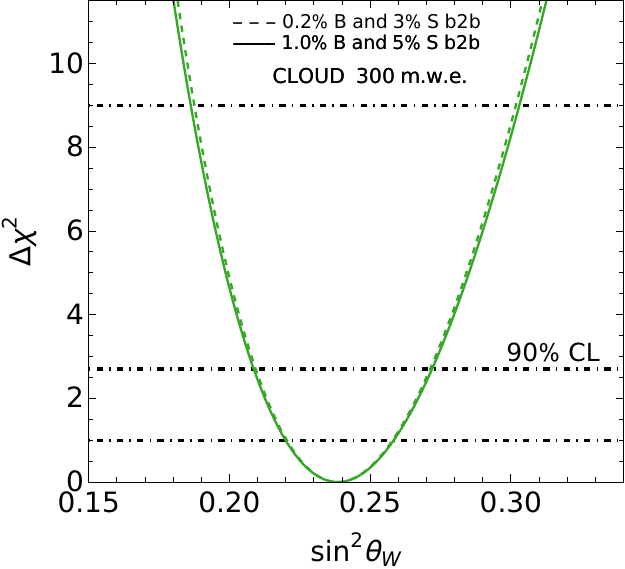}
		\end{subfigure}
		\hfill
		\begin{subfigure}[h]{0.495\textwidth}
			\caption{}
			\label{fb2bs2}
			\includegraphics[width=\textwidth]{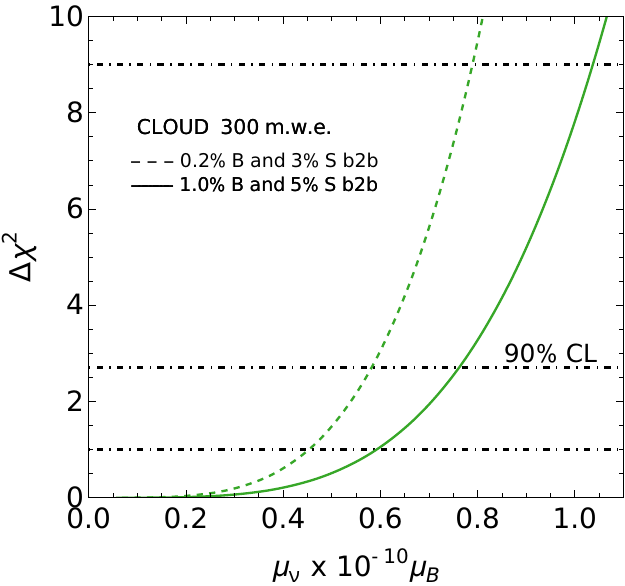}
		\end{subfigure}
		\hfill	
 \caption{The expected sensitivities ($\Delta \chi^2$) at the CLOUD setup. The left panel displays the sensitivity to the weak mixing angle ($\sin^2 \theta_W$), while the right panel shows the sensitivity to the effective neutrino magnetic moment ($\mu_\nu$). The solid lines consider the case of conservative systematic assumptions while dashed lines represent the case of optimistic systematics. See the text for a detailed explanation.}
  \label{fig:b2bcloud}
\end{figure}

In Figure~\ref{fig:b2bcloud}, we present the projected sensitivities ($\Delta \chi^2$) for the CLOUD setup. The left panel shows the sensitivity to the weak mixing angle ($\sin^2 \theta_W$), while the right panel shows the sensitivity to the effective neutrino magnetic moment ($\mu_\nu$). The solid lines correspond to conservative systematic assumptions, specifically a 5\% signal shape (S) and a 1\% background shape (B) uncertainty. The dashed lines consider an optimistic scenario, namely a 3\% signal shape and a 0.2\% background shape uncertainty. Thus, in the conservative scenario, the expected 1$\sigma$ CL sensitivity to the weak mixing angle at CLOUD is
\begin{equation}
\label{wmacloud}
\sin^2 \theta_{\mathrm{W}} = 0.239 \pm 0.019 ~(1\sigma)\,,
\end{equation}
representing a determination of the weak mixing angle with 8\% precision. For comparison, TEXONO reported $\sin^2\theta_W = 0.251 \pm 0.031\,(\mathrm{stat}) \pm 0.024\,(\mathrm{syst})$ from reactor $\bar{\nu}_e$--$e$ scattering~\cite{TEXONO:2009knm}, corresponding to an overall precision of about 16\%; therefore, the CLOUD conservative projection improves the precision by approximately a factor of two. On the other hand, the projected 90\% CL sensitivity to the effective neutrino magnetic moment ($\mu_{\nu}$) in the conservative scenario is
\begin{equation}
\label{nmmcloud}
\mu_{\nu} < 0.77 \times 10^{-10} \,\mu_{\mathrm{B}} \;.
\end{equation}
Hence, in this case, CLOUD sensitivity to the effective neutrino magnetic moment is comparable with the limit obtained by the TEXONO collaboration, $\mu_{\nu} < 0.74 \times 10^{-10} \,\mu_{\mathrm{B}}$~\cite{TEXONO:2006xds}, and improves upon the bound set by the MUNU collaboration, $\mu_{\nu} < 0.90 \times 10^{-10} \,\mu_{\mathrm{B}}$~\cite{MUNU:2005xnz}. However, the sensitivity at CLOUD remains weaker than the limit set by the GEMMA collaboration, $\mu_{\nu} < 0.29 \times 10^{-10} \,\mu_{\mathrm{B}}$~\cite{Beda:2012zz}.
\begin{figure}[H]
\begin{subfigure}[h]{0.495\textwidth}
			\caption{}
			\label{fbs1}
\includegraphics[width=\textwidth]{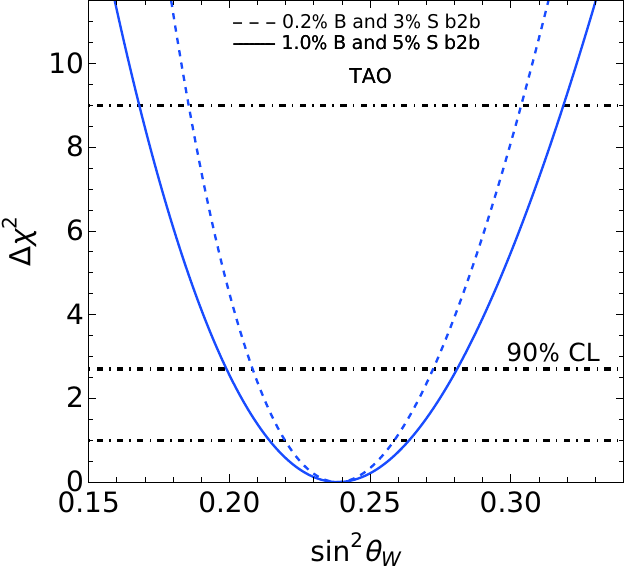}
		\end{subfigure}
		\hfill
		\begin{subfigure}[h]{0.495\textwidth}
			\caption{}
			\label{fbs2}
			\includegraphics[width=\textwidth]{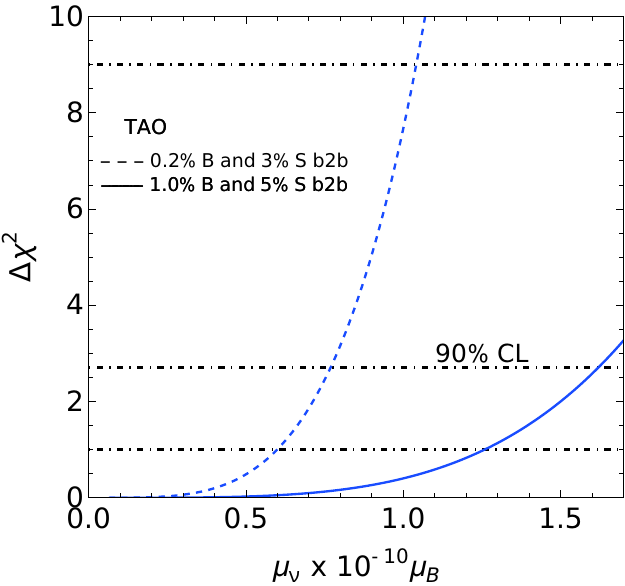}
		\end{subfigure}
		\hfill	
 \caption{The projected sensitivities ($\Delta \chi^2$) at the TAO configuration. The left panel shows the sensitivity to the weak mixing angle ($\sin^2 \theta_W$), while the right panel displays the sensitivity to the effective neutrino magnetic moment ($\mu_\nu$). The solid lines consider the case of conservative systematic assumptions while dashed lines represent the case of optimistic systematics. See the text for a detailed explanation.}
  \label{fig:b2btao}
\end{figure}
In Figure~\ref{fig:b2btao}, we display the projected sensitivities ($\Delta \chi^2$) for the TAO configuration. The sensitivity to the weak mixing angle ($\sin^2 \theta_W$) is shown in the left panel, and the sensitivity to the effective neutrino magnetic moment ($\mu_\nu$) is shown in the right panel. The solid lines consider our conservative systematics scenario: $\sigma_S =  5\%$ for signal events and $\sigma_B =  1\%$ for background contaminants. Besides, dashed lines assume an optimistic scenario: 3\% signal shape and 0.2\% background shape, accordingly.

Hence, in the conservative scenario, the expected 1$\sigma$ CL sensitivity to the weak mixing angle at TAO is
\begin{equation}
\label{wmatao}
\sin^2 \theta_W =0.239^{+0.026}_{-0.024}~(1\sigma)\,,
\end{equation}
thus, TAO could provide a determination of the weak mixing angle at the 11\% level precision. In addition, within the conservative systematics case, the projected 90\% CL sensitivity to the effective neutrino magnetic moment ($\mu_{\nu}$) is
\begin{equation}
\label{nmmtao}
\mu_{\nu} < 1.63 \times 10^{-10} \mu_{\rm B} \,.
\end{equation}
Therefore, in this scenario, the expected sensitivity to $\mu_{\nu}$ at TAO is comparable with the limit obtained by ROVNO, $\mu_{\nu} < 1.9 \times 10^{-10} \,\mu_{\mathrm{B}}$~\cite{Derbin:1993wy}, and improves upon the bound set by KRASNOYARSK, $\mu_{\nu} < 2.4 \times 10^{-10} \,\mu_{\mathrm{B}}$~\cite{Vidyakin:1992nf}.
\begin{figure}[H]
\begin{subfigure}[h]{0.492\textwidth}
			\caption{}
			\label{fs1}
\includegraphics[width=\textwidth]{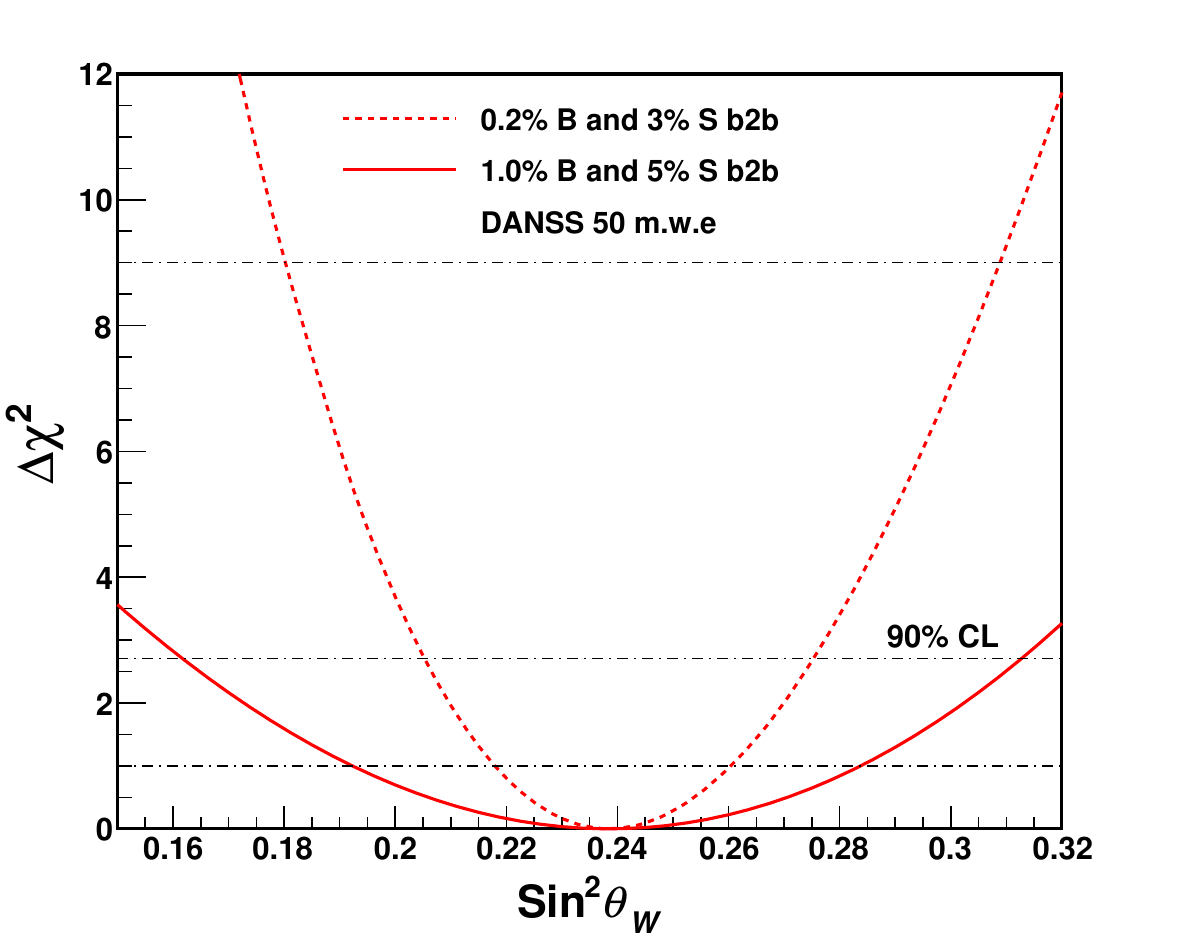}
		\end{subfigure}
		\hfill
		\begin{subfigure}[h]{0.497\textwidth}
			\caption{}
			\label{fs2}
			\includegraphics[width=\textwidth]{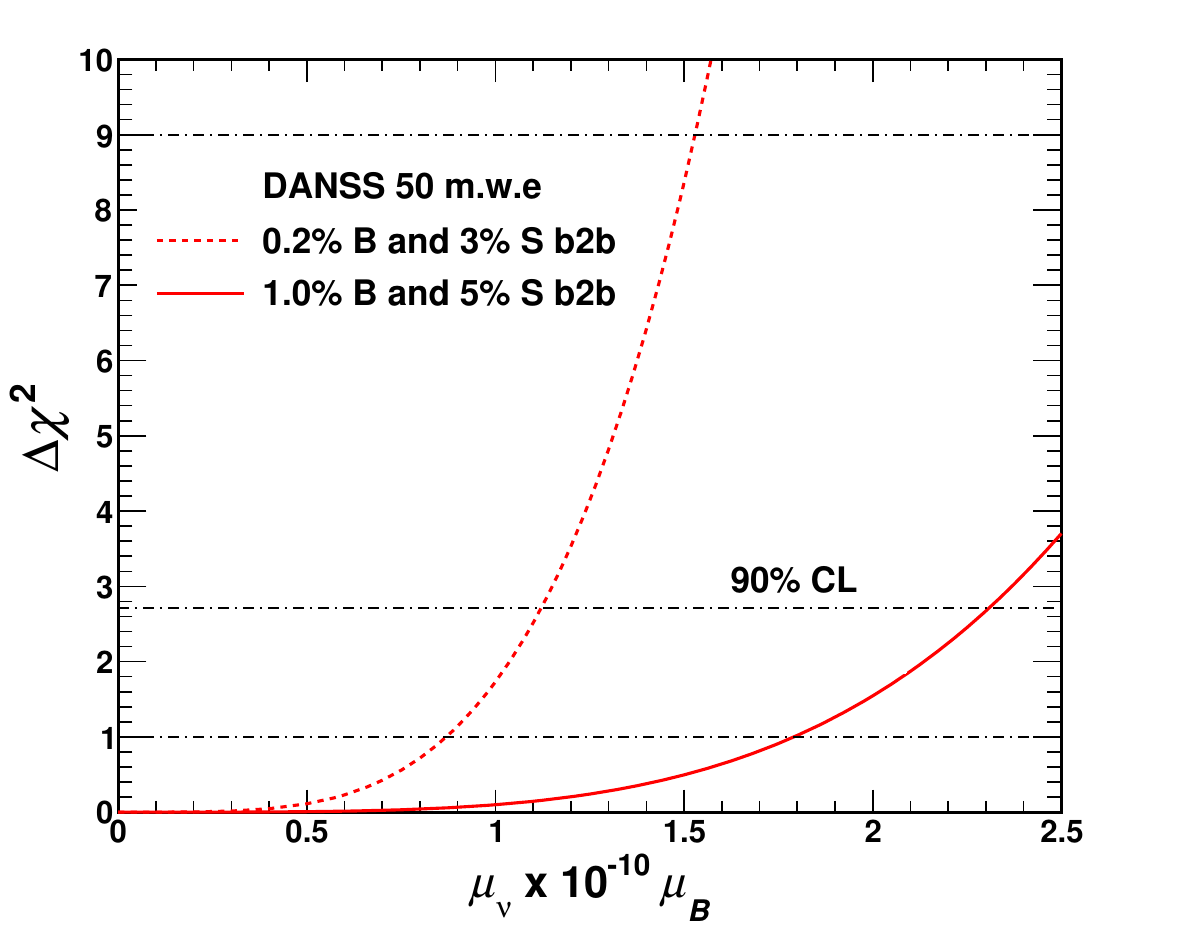}
		\end{subfigure}
		\hfill	
 \caption{The expected sensitivities ($\Delta \chi^2$) at the DANSS setup. The left panel displays the sensitivity to the weak mixing angle ($\sin^2 \theta_W$), while the right panel shows the sensitivity to the effective neutrino magnetic moment ($\mu_\nu$). The solid lines consider the case of conservative systematic assumptions while dashed lines represent the case of optimistic systematics. See the text for a detailed explanation.}
  \label{fig:b2bdanss}
\end{figure}
In Figure~\ref{fig:b2bdanss}, we present the projected sensitivities ($\Delta \chi^2$) for the DANSS setup. The sensitivity to the weak mixing angle ($\sin^2 \theta_W$) is shown in the left panel, and the sensitivity to the effective neutrino magnetic moment ($\mu_\nu$) is shown in the right panel. As in the case of CLOUD and TAO, here, solid lines assume conservative systematics, while dashed lines represent the optimistic scenario of systematic uncertainties. As displayed in Fig.~\ref{fig:b2bdanss}, within the conservative scenario, the expected 1$\sigma$ CL sensitivity to the weak mixing angle at DANSS is
\begin{equation}
\label{wmadans}
\sin^2 \theta_W =0.239^{+0.044}_{-0.047}~(1\sigma)\,,
\end{equation}
which represents a determination of the weak mixing angle at the 20\% level precision. Besides, the projected 90\% CL sensitivity to the effective neutrino magnetic moment ($\mu_{\nu}$) in the conservative scenario is
\begin{equation}
\label{nmmdans}
\mu_{\nu} < 2.32 \times 10^{-10} \mu_{\rm B} \,.
\end{equation}
In this case, DANSS sensitivity is comparable to the limit set by KRASNOYARSK, $\mu_{\nu} < 2.4 \times 10^{-10} \,\mu_{\mathrm{B}}$~\cite{Vidyakin:1992nf}. 

A summary of the corresponding sensitivities for each configuration is shown in Table~\ref{tab:comp}.
The impact of systematic uncertainties on the sensitivity is considerable.
Comparing the optimistic ($\sigma_S = 3\%$, $\sigma_B = 0.2\%$) and conservative ($\sigma_S = 5\%$, $\sigma_B = 1\%$) scenarios, the precision of $\sin^2\theta_W$ improves by approximately 22\% for TAO and 54\% for DANSS, while no significant improvement is observed for CLOUD, whose sensitivity is already dominated by statistical uncertainties rather than systematics.
The reduction of the signal shape uncertainty from 5\% to 3\% alone improves the $\sin^2\theta_W$ precision by about 4\% for TAO and 2\% for DANSS, while the remaining gain comes from reducing $\sigma_B$ from 1\% to 0.2\%, which is the dominant contribution.
Regarding the effective neutrino magnetic moment $\mu_\nu$, the upper limit improves by approximately a factor of two for both TAO and DANSS under the optimistic scenario.
Notably, CLOUD provides the strongest constraints on $\mu_\nu$ in both the conservative and optimistic scenarios, owing to its excellent signal-to-background ratio.

\begin{table}[H]                           
      \centering  
      \begin{tabular}{lrrr}
      \hline \hline
          Expected sensitivity  &~~ CLOUD ~~&~~TAO ~~&~~DANSS \\  \hline
          \multicolumn{4}{c}{Conservative scenario ($\sigma_S = 5\%$, $\sigma_B = 1\%$)} \\  \hline           
          Weak mixing angle: $\sin^2 \theta_W \pm 1\sigma$      &~~$0.239^{+0.019}_{-0.019}$ &~~$0.239^{+0.026}_{-0.024}$           &~~$0.239^{+0.044}_{-0.047}$  \\  
          Magnetic moment $[\times 10^{-11} \mu_{\text{B}}]~$(90\% CL)       & $\mu_{\nu} < 7.7$    & $\mu_{\nu} < 16.3$  & $\mu_{\nu} 
  < 23.2$  \\ \hline                            
          \multicolumn{4}{c}{Optimistic scenario ($\sigma_S = 3\%$, $\sigma_B = 0.2\%$)} \\\hline  
          Weak mixing angle: $\sin^2 \theta_W \pm 1\sigma$      &~~$0.239^{+0.019}_{-0.019}$ &~~$0.239^{+0.020}_{-0.019}$ &~~$0.239^{+0.021}_{-0.021}$  \\                                   
          Magnetic moment $[\times 10^{-11} \mu_{\text{B}}]~$(90\% CL)       & $\mu_{\nu} < 5.9$    & $\mu_{\nu} < 7.8$  & $\mu_{\nu} < 11.5$  \\  \hline \hline
      \end{tabular}                                          
      \caption{The expected sensitivities at the reactor-based experiments for both the conservative and optimistic systematic         
  scenarios.}                                                                                                                          
      \label{tab:comp}                                                                                                                 
  \end{table}     

\begin{figure}[H]
\includegraphics[scale=0.7]{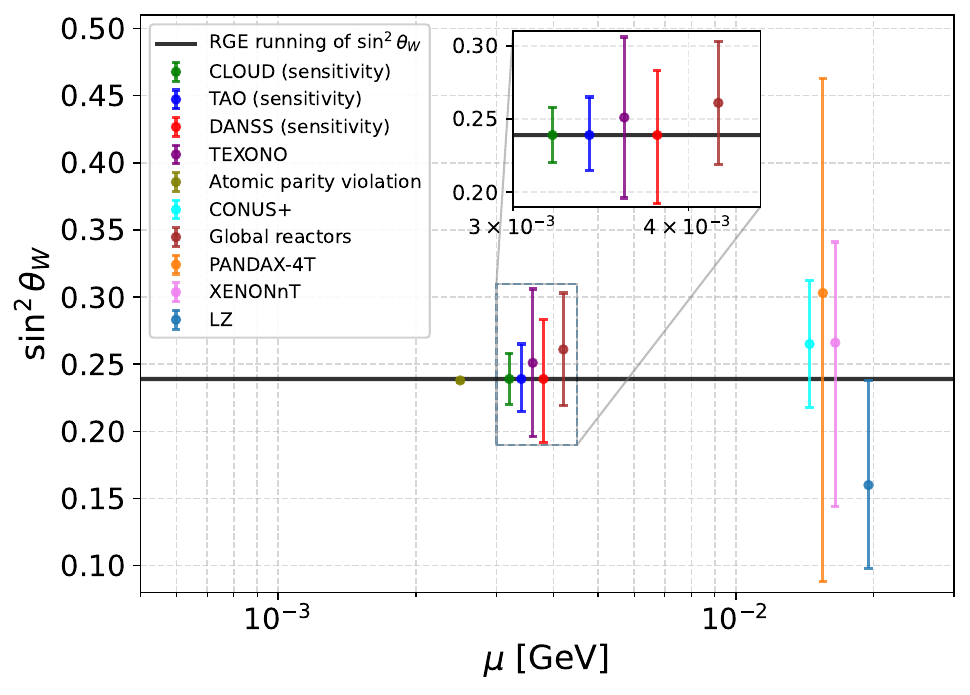}
		 \caption{Scale dependence $\mu$ [GeV] of $\sin^2 \theta_W$ in the $\overline{\mathrm{MS}}$ renormalization scheme~\cite{Erler:2017knj,ParticleDataGroup:2024cfk}. We present our expected 1$\sigma$ precision sensitivities of the weak mixing angle assuming the conservative systematic scenario~($\sigma_S =  5\%$, $\sigma_B =  1\%$) for CLOUD (green), TAO (blue), and DANSS (red). For comparison, we also show the results from TEXONO~\cite{TEXONO:2009knm} (purple), global fit reactors~\cite{Canas:2016vxp} (maroon), Atomic Parity Violation (APV)~\cite{Antypas:2018mxf} (olive), as well as CONUS$+$~\cite{Ackermann:2025obx, Alpizar-Venegas:2025wor} (cyan) via coherent elastic neutrino-nucleus scattering (CE$\nu$NS). Furthermore, we also display the results from solar neutrino interactions via CE$\nu$NS processes in dark matter direct detection experiments: PANDAX-4T, XENONnT, and LUX-ZEPLIN (LZ)~\cite{Maity:2024aji, DeRomeri:2024iaw, LZ:2025igz}. See the text for details.} 
  \label{fig:rge}
\end{figure}
Fig.~\ref{fig:rge} shows the evolution of the weak mixing angle as a function of the energy scale ($\mu$) within the $\overline{\mathrm{MS}}$ renormalization scheme. We display our projected 1$\sigma$ precision for the weak mixing angle at CLOUD (green), TAO (blue), and DANSS (red). For comparison, we also include results from other $\sin^2\theta_W$ measurements, TEXONO~\cite{TEXONO:2009knm} (purple), APV~\cite{Antypas:2018mxf} (olive) , a global fit of reactor data (TEXONO, MUNU, ROVNO, and KRASNOYARSK)~\cite{Canas:2016vxp} (maroon), as well as recent assessments from CONUS$+$ experiment (cyan) obtained via coherent elastic neutrino-nucleus scattering~\cite{Ackermann:2025obx, Alpizar-Venegas:2025wor}. We further present the results of solar neutrino analyses from CE$\nu$NS interactions for the PandaX-4T (orange), XENONnT (pink), and LUX-ZEPLIN (light-blue) dark matter direct detection experiments~\cite{Maity:2024aji, DeRomeri:2024iaw, LZ:2025igz}. For instance, as compared to the global reactor fit, the projected sensitivities for CLOUD and TAO may significantly improve the precision determination of the weak mixing angle. Furthermore, the DANSS experiment is projected to yield a more precise measurement than that obtained from TEXONO.

\subsection{Transition magnetic moments}
In previous subsections, we assessed the sensitivities to the weak mixing angle, $\sin^2 \theta_W$, and the effective neutrino magnetic moment, $\mu_\nu$, at the aforementioned reactor neutrino experiments. Here, we present our results on the projected sensitivities to the neutrino transition magnetic moments.

\begin{figure}[H]
\includegraphics[scale=0.6]{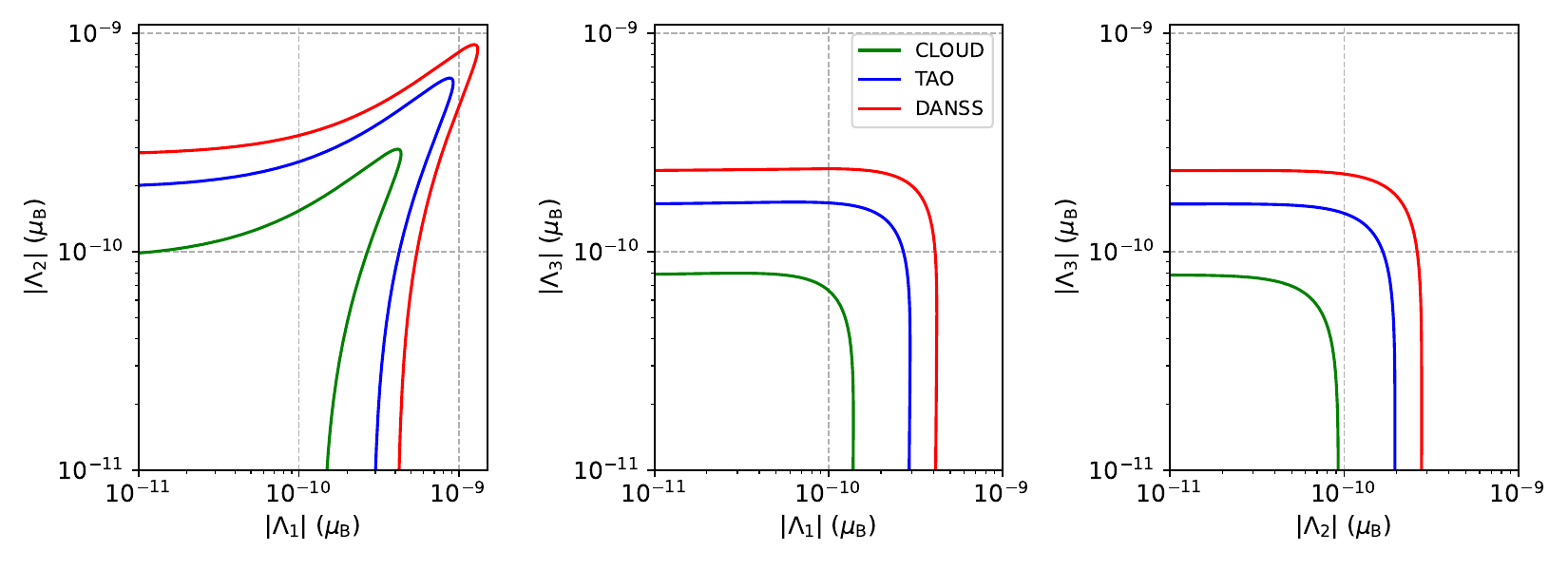}
		 \caption{The expected 90\% CL limits on the neutrino transition magnetic moments, $\abs{\Lambda_j}$, at CLOUD (green), TAO (blue), and DANSS (red) configuration, assuming the conservative systematic scenario ($\sigma_S = 5\%$, $\sigma_B = 1\%$). Except for $\delta_{CP}$, we consider the best fit oscillation parameters from Ref.~\cite{Esteban:2024eli}. See the text for a detailed explanation.} 
  \label{fig:tmm}
\end{figure}
In fig.~\ref{fig:tmm}, we present the expected 90\% CL limits on the neutrino transition magnetic moments $\abs{\Lambda_j}$ at the CLOUD (green), TAO (blue), and DANSS (red) setup, accordingly. To obtain the corresponding transition magnetic moments, in Eq.~\ref{eq:mueff}, for each experimental configuration, we consider the projected 90\% CL bound on the effective neutrino magnetic moment ($\mu_\nu$) as given in Table~\ref{tab:comp}, and assessed the impact of one parameter at a time, fixing all the remaining transition magnetic moment parameters and phases ($\omega_j$) to zero (including $\delta_{CP}$), see, e.g. Ref.~\cite{Canas:2015yoa}. Aside from $\delta_{CP}$, we fix other relevant oscillation parameters to their best fit value from Ref.~\cite{Esteban:2024eli}. Our expected limits on the transition magnetic moments $\abs{\Lambda_j}$, along with comparisons to other bounds from reactor-based experiments, are summarized in Table~\ref{tab:TMM}. Besides, in the optimistic scenario ($\sigma_S = 3\%$, $\sigma_B = 0.2\%$), the limits on the neutrino transition magnetic moments $|\Lambda_j|$ could improve by approximately a factor of two.

\begin{table}[H]
\caption{90\% CL limits on the neutrino transition magnetic moments in the mass basis, $\abs{\Lambda_j}$, from reactor neutrino experiments. The expected sensitivities at the reactor-based experiments considered in this study, assuming the conservative systematic scenario~($\sigma_S =  5\%$, $\sigma_B = 1\%$.}
\begin{tabular}{l p{12mm} c p{1cm} c p{1cm} c}
\hline
Experiment & & $|\Lambda_{1}|$ &&  $|\Lambda_{2}|$ & & $|\Lambda_{3}|$\\ [0.5ex]
\hline
CLOUD (this work) & & $ 1.5 \times 10^{-10}\mu_{B}$&& $0.9\times10^{-10}\mu_{B}$& & $0.8\times10^{-10}\mu_{B}$ \\
TAO (this work) & & $ 3.0 \times 10^{-10}\mu_{B}$&& $2.0\times10^{-10}\mu_{B}$& & $1.7\times10^{-10}\mu_{B}$ \\
DANSS (this work) & & $ 4.0 \times 10^{-10}\mu_{B}$&& $3.0\times10^{-10}\mu_{B}$& & $2.5\times10^{-10}\mu_{B}$ \\
KRASNOYARSK~\cite{Canas:2015yoa} & & $ 4.7\times10^{-10}\mu_{B}$& & $3.3\times10^{-10}\mu_{B}$& & $2.8\times10^{-10}\mu_{B}$\\
ROVNO~\cite{Canas:2015yoa} & & $ 3.0 \times 10^{-10}\mu_{B}$&& $2.1\times10^{-10}\mu_{B}$& & $1.8\times10^{-10}\mu_{B}$ \\
MUNU~\cite{Canas:2015yoa} & & $2.1 \times 10^{-10}\mu_{B}$&& $1.5\times10^{-10}\mu_{B}$& & $1.3\times10^{-10}\mu_{B}$ \\
TEXONO~\cite{Canas:2015yoa} & & $ 3.4 \times 10^{-10}\mu_{B}$ & &$2.4\times10^{-10}\mu_{B}$& & $2.0\times10^{-10}\mu_{B}$ \\
GEMMA~\cite{Canas:2015yoa} & & $ 5.0 \times 10^{-11}\mu_{B}$ & &$3.5\times10^{-11}\mu_{B}$& & $2.9\times10^{-11}\mu_{B}$ \\ 
\hline
\end{tabular}
\label{tab:TMM}
\end{table}

\subsection{Non-standard neutrino interactions}
By neglecting flavor-changing couplings, limits on either $\varepsilon_{ee}^{e R}$ or $\varepsilon_{ee}^{e L}$ can be derived by setting the other to zero.
Under the conservative systematic scenario ($\sigma_S = 5\%$, $\sigma_B = 1\%$), the expected constraints on these NSI couplings at reactor experiments are summarized in Table~\ref{tab:nsi_constraint}. The projected constraints on NSI parameters from both CLOUD and TAO surpass the existing limit from the TEXONO experiment~\cite{TEXONO:2010tnr}. While DANSS is expected to set relatively looser bounds, they remain comparable to the current TEXONO result.

\begin{table}[H]
    \centering
    \begin{tabular}{lrrr}
    \hline \hline
        NSI coupling constraint  &~~ $\varepsilon_{ee}^{e R}$ ~~&~~ $\varepsilon_{ee}^{e L}$  \\  \hline
        CLOUD      &~~$(-0.03,       0.03)$ &~~$(-0.60, 0.56)$   \\
        TAO      &~~$(-0.04,       0.04)$ &~~$(-0.65, 0.70)$   \\
        DANSS      &~~$(-0.07,       0.07)$ &~~$(-0.97, 0.99)$   \\
        TEXONO~\cite{TEXONO:2010tnr}      &~~$(-0.07,0.08)$ &~~$(-1.53, 0.38)$   \\
        \hline \hline
    \end{tabular}
    \caption{The expected 90\% CL constraints on the non-standard neutrino interaction couplings at the reactor-based experiments, assuming the conservative systematic scenario ($\sigma_S = 5\%$, $\sigma_B = 1\%$).}
    \label{tab:nsi_constraint}
\end{table}
\section{Conclusions}
\label{sec:conclusions}

In this paper we explored the physics potential of neutrino-electron elastic scattering at current and future reactor neutrino experiments. We compute the expected sensitivities to the weak mixing angle, effective neutrino magnetic moment, and the transition magnetic moments at the CLOUD, TAO, and DANSS configurations. Regarding the weak mixing angle, even under conservative assumptions of systematic uncertainties (Table~\ref{tab:sys}), the projected sensitivities for CLOUD and TAO may significantly improve the precision determination of $\sin^2 \theta_W$ compared to the global fit of reactor datasets (Fig.~\ref{fig:rge}). Furthermore, as displayed in Table~\ref{tab:comp}, the DANSS experiment is expected to improve upon the TEXONO measurement (Fig.~\ref{fig:rge}). Regarding the effective magnetic moment, for the assumed energy threshold in this analysis ($T_{\rm{th}} = 1$ MeV), as presented in Fig.~\ref{fig:b2bcloud}, Fig.~\ref{fig:b2btao}, and Fig.~\ref{fig:b2bdanss}, the projected sensitivities to the effective neutrino magnetic moment, $\mu_\nu$, were found to be comparable with previous measurements (Sec.~\ref{sec:results}). In addition, based on the expected sensitivities to the effective neutrino magnetic moment (Fig.~\ref{fig:tmm}), we compute the corresponding limits on the neutrino transition magnetic moments; our results on this assessment are summarized in Table~\ref{tab:TMM}.

Finally, within the conservative scenario studied here, the reactor-based experiments CLOUD, TAO, and DANSS are projected to place competitive constraints on the left- and right-handed NSI couplings, $\varepsilon_{ee}^{e L}$ and $\varepsilon_{ee}^{e R}$. As summarized in Table~\ref{tab:nsi_constraint}, CLOUD achieves the tightest constraints. TAO provides a sensitivity on $\varepsilon_{ee}^{e R}$ that closely approaches CLOUD's, while its constraint on $\varepsilon_{ee}^{e L}$ is slightly weaker. DANSS yields the least stringent bounds, which nonetheless remain comparable to the current best limit. These results demonstrate the strong potential of near-future reactor experiments to probe and constrain non-standard neutrino interactions.

Nevertheless, for all setups, an improved background model would enhance the sensitivity to the weak mixing angle, the effective neutrino magnetic moment, and neutrino transition magnetic moments. Specific needs vary by experiment: a more precise energy response model is crucial for CLOUD, while for TAO, effects such as liquid scintillator nonlinearities will grow in importance as background rates decrease. In the case of DANSS, analyzing reactor-off data from its ongoing IBD measurements would help mitigate key background contaminants. Furthermore, lowering the energy threshold across these configurations would significantly boost the sensitivity to the effective neutrino magnetic moment.

\section*{Acknowledgments}
We acknowledge useful discussions with Vitalii Zavadaskyi, Maxim Gonchar, Guofu Cao, Yufeng Li, Omar Miranda, Anatael Cabrera and Hiroshi Nunokawa. This work is supported in part by the National Natural Science Foundation of China under grant No.12342502. We thank the anonymous referee for the comments and suggestions that have helped us to improve our manuscript.

This paper represents the views of the authors and should not be considered a CLOUD, JUNO or DANSS collaboration paper.

\end{document}